\documentclass[aps,prl,twocolumn,groupedaddress,showpacs]{revtex4}

\usepackage{graphicx}
\usepackage{dcolumn}
\usepackage{bm}

\begin{document}

\title{Cold atom gas at very high densities in an optical surface microtrap}

\author{M. Hammes}
\author{D. Rychtarik}
\author{H.-C. N\"agerl}
\author{R. Grimm}

\affiliation{Institut f\"ur Experimentalphysik, Universit\"at
Innsbruck, Technikerstr.\ 25, A-6020 Innsbruck, Austria}

\date{\today}

\begin{abstract}
An optical microtrap is realized on a dielectric surface by
crossing a tightly focused laser beam with an horizontal
evanescent-wave atom mirror. The nondissipative trap is loaded
with $\sim$$10^5$ cesium atoms through elastic collisions from a
cold reservoir provided by a large-volume optical surface trap.
With an observed 300-fold local increase of the atomic number
density approaching $10^{14}{\rm cm}^{-3}$, unprecedented
conditions of cold atoms close to a surface are realized.
\end{abstract}


\maketitle

The trapping of cold atoms very close to material objects has
attracted considerable interest \cite{Ovchinnikov1997a,
Gauck1998a, Schneble1999a, Reichel1999a, Muller1999a, Dekker2000a,
Folman2000a, ott2001a, hansel2001b, leanhardt2002a} and opens up
fascinating perspectives for a great variety of experiments
ranging from fundamental studies of surface interactions to the
realization of low-dimensional quantum gases to applications in
the field of quantum information processing. Material structures
that carry currents, produce electric fields, or transport light
are the key to realize novel trapping configurations such as
microtraps and integrated devices for atom optics. In this context
a surface constitutes an interface that connects the atomic
quantum system to the environment with the prospect of gaining
control and enabling measurements on the quantum level.

For experiments in this research field, various surface trapping
and guiding schemes have been devised: Magneto-optical traps near
surfaces \cite{Ovchinnikov1997a, Schneble1999a, Reichel1999a},
magnetic surface traps and guides \cite{Reichel1999a, Muller1999a,
Dekker2000a, Folman2000a}, and optical dipole traps and guides
\cite{Ovchinnikov1997a, Gauck1998a} represent the three main
classes. Very recently magnetic surface traps could be combined
with Bose-Einstein condensates (BEC) \cite{ott2001a, hansel2001b,
leanhardt2002a}. While in present magnetic surface trapping
schemes the atomic samples are still separated from the surface by
typically a few ten $\mu$m, optical schemes
\cite{Ovchinnikov1997a, Gauck1998a, Hammes2000a, Hammes2001a} have
already trapped atoms much closer to the surface with distances on
the order of the optical wavelength and quasi-2D conditions have
been attained in low-density samples. Optical surface trapping
fields thus offer an intriguing potential for future nanotraps and
guides based on integrated optical waveguides \cite{burke2002a}.

\begin{figure}
\includegraphics[width=7.5cm]{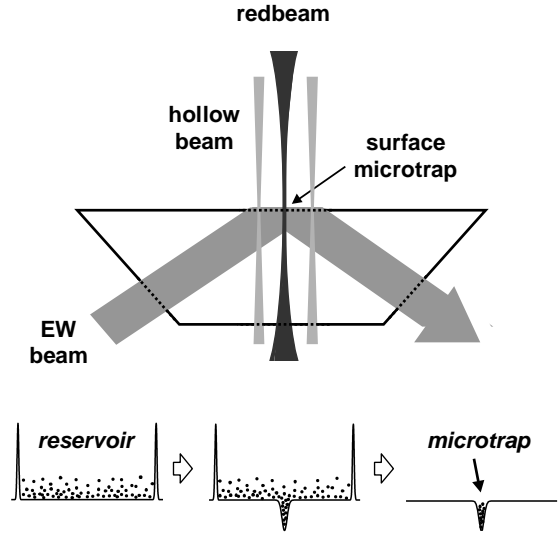}
\caption{\label{trapscheme} Scheme of the surface trap together
with an illustration of collisional microtrap loading, in which
the horizontal trapping potential is changed in three steps.}
\end{figure}

In this Letter, we demonstrate a novel optical surface trap based
on an evanescent-wave (EW) atom mirror in combination with a
tightly focused red-detuned laser beam; see Fig.~\ref{trapscheme}.
This optical dipole trap confines atoms in a very small volume
very close to a dielectric surface. Efficient loading through
elastic collisions allows us to trap a large number of atoms from
an optically precooled reservoir and thus to reach exceptionally
high densities even with a classical gas. This scheme also opens
up a possible new route for an all-optical production of a
Bose-Einstein condensate \cite{Barrett2001a} with prospects for
future experiments on quantum-degenerate gases near the surface.


The starting point for our experiments with cesium atoms is the
``gravito-optical surface trap'' \cite{Ovchinnikov1997a,
Hammes2000a, Hammes2001a}. This trap is realized under ultrahigh
vacuum conditions ($p \lesssim 10^{-11}\,$mbar) by combining the
EW atom mirror with a blue-detuned hollow beam (HB). The EW is
produced with the 55-mW beam of a diode laser (SDL-5712-H1) in a
nearly round spot with a 1/$e^2$-radius of 0.65\,mm on the surface
of a fused-silica prism. The angle of incidence is about
$2^{\circ}$ above the critical angle of total internal reflection
and the polarization lies in the plane of incidence. A heated
cesium cell is used to filter resonant background light out of
this beam \cite{Hammes2001a}. The HB for horizontal confinement is
derived from a Ti:Sapphire laser (Coherent 899-01/895) at a
wavelength of 849\,nm, i.e.\ with a detuning of about $\sim$3\,nm
from the 852-nm Cs resonance line. It is shaped with a special
axicon optics \cite{Manek1998a} to provide a two-dimensional
cylindrical box potential. The HB has a total power of 330\,mW, a
diameter of 820\,$\mu$m, and a potential height on the order of
100\,$\mu$K. Typically $2 \times 10^7$ atoms are initially loaded
from a magneto-optical trap (MOT) operating a few mm above the
surface. First, the optical detuning of the EW is set to 5\,GHz to
implement Sisyphus cooling by inelastic reflections
\cite{Ovchinnikov1997a}. After 5\,s of optical cooling the EW
detuning $\Delta \nu_{\rm EW}$ is increased in a 2-s linear ramp
to a large value in the range between 100\,GHz and 200\,GHz; this
frequency ramp is accomplished by rapid temperature tuning of the
laser diode. In this way a conservative trapping potential is
realized with negligible photon scattering and heating below
100\,nK/s. The large detuning also strongly reduces light-induced
collisional loss in the blue-detuned trap light
\cite{Hammes2000a}.

Our surface reservoir prepared in this way contains $2 \times
10^6$ Cs atoms at a temperature of $T \approx 3\,\mu$K at $\Delta
\nu_{\rm EW} = 160\,$GHz. The vertical density distribution
follows the barometric equation with a $1/e$-height of
$\sim$$20\mu$m \cite{Ovchinnikov1997a} with a peak value close to
the EW atom mirror of $2 \times 10^{11}$cm$^{-3}$. The peak
phase-space density of the unpolarized sample in the $F=3$ ground
state is $\sim$\,$2 \times 10^{-5}$, where we assume that the
sample at low magnetic field ($\lesssim 5\,\mu$T) is equally
distributed over the seven magnetic substates. The elastic
collision rate is of the order of 20\,s$^{-1}$, leading to a
thermalization time of $\sim$0.5\,s. At the large EW detuning, the
lifetime of the reservoir is limited to typically 1.5\,s. We
believe that this is due to escape through weaker regions of the
EW potential barrier in the Gaussian profile and at local surface
defects.

All measurements reported here are based on recapturing the atoms
trapped on the EW atom mirror into the MOT. The number of atoms is
determined from the integrated fluorescence detected with a
carefully calibrated CCD camera. For the absolute atom number we
estimate a calibration error below a factor of 1.5. The
temperature is measured with a release-and-recapture method by
turning off the EW for a short time interval and measuring the
relative loss of atoms \cite{Ovchinnikov1997a}.

For implementing the surface microtrap we use the beam of a Nd:YAG
laser at a wavelength of 1064\,nm, which is focused right into the
center of the cesium ensemble with a power of 330\,mW and a waist
of $w_0 = 32\,\mu$m. There it produces a Gaussian-shaped
horizontally attractive optical potential with a calculated depth
of a $\hat{U}/k_B = 48\,\mu$K \cite{Grimm2000a}. In a harmonic
approximation to the trap center the oscillation frequency is
$\omega_{\rm r} = (4\hat{U}/m w^2_0)^{1/2} = 2\pi \times 540$\,Hz.
After turning on the Nd:YAG beam, this additional narrow potential
well is filled by thermalizing elastic collisions. Since the
narrow well is deep compared with the temperature of the reservoir
($\hat{U}/k_B T \approx 15$) a very large increase of the local
density can be expected. After loading the Nd:YAG beam the
reservoir can be removed by shutting down the HB. In this way, a
very dense sample of atoms stored in the surface microtrap is
realized.

This highly efficient collisional loading scheme plays a crucial
role in our experiments as it leads to a drastic local increase of
number and phase-space density. The basic idea of a local gain by
adiabatically changing the potential shape was first pointed out
in \cite{Pinkse1997a}. Our scheme is similar to the ``dimple
trap'' desribed in \cite{Stamper-Kurn1998c} where an optical
dipole trap was combined with a magnetic trap and a 50-fold
increase in local phase-space density was reached. We believe that
this mechanisum also plays an important role for trap loading in
the all-optical BEC experiment of Ref.~\onlinecite{Barrett2001a},
where two tightly focused CO$_2$-laser beams formed sort of a
dimple in their crossing region. The efficient ``dimple trick''
may also find very interesting applications in context with
degenerate Fermi gases as suggested in \cite{Viverit2001a}.

\begin{figure}
\includegraphics[width=7cm]{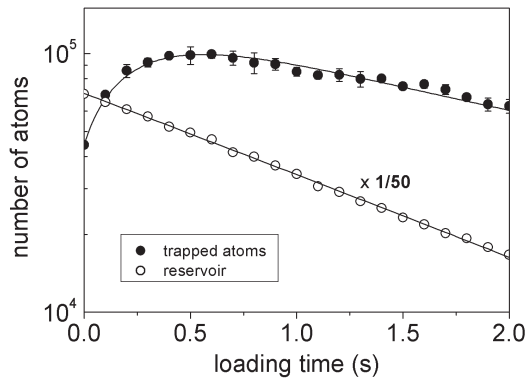}
\caption{\label{loading} Temporal evolution of the number of Cs
atoms in the reservoir ($\circ$) and in the microtrap ($\bullet$)
together with fits (solid lines) according to the simple model
described in the text.}
\end{figure}

In order to study the collisional loading process we suddenly turn
on the Nd:YAG beam after preparation of the reservoir with $\Delta
\nu_{\rm EW} = 120$\,GHz and $T \approx 4\mu$K. The measured
number of atoms loaded into the microtrap with increasing time is
displayed in Fig.~\ref{loading} together with the number of
reservoir atoms. The latter is very well described by an
exponential decay $N_{\rm res} = N_{\rm res, 0} \exp(-t/\tau_{\rm
res})$ with $N_{\rm res, 0} = 3.5 \times 10^6$ and $\tau_{\rm res}
= 1.4\,$s. At zero loading time, a sudden transfer puts an initial
number of $4 \times 10^4$ atoms into the phase-space of the
microtrap. This transfer of $\sim$1\% agrees very well with a
calculation of the sudden transfer ratio. Thermalizing elastic
collisions then fill the optical microtrap on the timescale of a
few 100\,ms and the trapped number of atoms reaches a maximum of
$\sim$$10^5$ after about 500\,ms. The atomic ensemble in the
microtrap then decays slower than the reservoir, which is due to
the high EW potential barrier in the middle of the reservoir.

We can model the loading in a simple way by a differential
equation for the number $N(t)$ of atoms in the microtrap,
\begin{equation}
\dot{N} = - \kappa N_{\rm res} (N - a N_{\rm res}) - b N^2 .
\end{equation}
The loading model is based on the experimental facts that the
temperature stays constant and that the microtrap does not
significantly affect the reservoir. By introducing the parameter
$a \ll 1$ we assume that thermalization tends to load a certain
fraction of the reservoir atoms into the microtrap. The loading
rate $\kappa N_{\rm res}$ is assumed to be proportional to the
elastic collision rate in the reservoir. For the loss out of the
microtrap, we consider two-body collisions as the main mechanism.
The loss coefficient is determined independently to $b = 7.2
\times 10^{-6}$\,s$^{-1}$ by observing the decay in the absence of
a reservoir ($N_{\rm res}=0$). A fit of $N(t)$ to the experimental
loading data in Fig.~\ref{loading} then yields $\kappa =
3.3\times10^{-7}$\,s$^{-1}$ and $a = 0.08$. The resulting initial
loading rate $\kappa N_{\rm res, 0} \approx 1\,$s$^{-1}$ is of the
same order as thermalization rate of the reservoir. The parameter
$a \approx 8$\,\% indicates the thermal equilibrium transfer ratio
into the microtrap, which exceeds the sudden transfer by almost an
order of magnitude.

The loaded atom number can be maximized by turning on the Nd:YAG
beam right after the optical cooling phase at the begin of the 2-s
EW detuning ramp. When right after the ramp the HB is turned off
and the reservoir atoms completely disappear after 200ms, we
observe up to $N=1.5\times10^5$ atoms in the microtrap.

The density distribution can be calculated under the assumption of
thermal equilibrium from the potential shape and the measured
temperature. The trap shape with its cylindrical symmetry is
approximately harmonic in the two dimensions of the horizontal
$x,y$-plane and has a one-dimensional wedge shape in the vertical
$z$-direction. In such a potential the density distribution can be
described by
\begin{equation}
n({\bf r}) = n_0 \exp(-\rho^2/\rho^2_0) \exp(-z/z_0)
\label{density}
\end{equation}
where $\rho^2=x^2+y^2$. The $1/e$-decrease of the density is
characterized by a horizontal radius $\rho_0=\omega_{\rm
r}^{-1}\sqrt{2k_BT/m}$ and a height $z_0=k_BT/mg$; here $m$
denotes the mass of a Cs atom and $g$ represents the gravitational
acceleration. With an effective volume defined as $V_0 = \pi
\rho^2_0 z_0$, the peak number density $n_0$ is related to the
trapped atom number and the effective volume by $n_0 = N/V_0$.
Note that the mean density and the mean quadratic density in this
trapping potential are related to the peak density by $\langle n
\rangle = n_0/4$ and $\langle n^2 \rangle = n_0^2/9$.

\begin{figure}
\includegraphics[width=8.5cm]{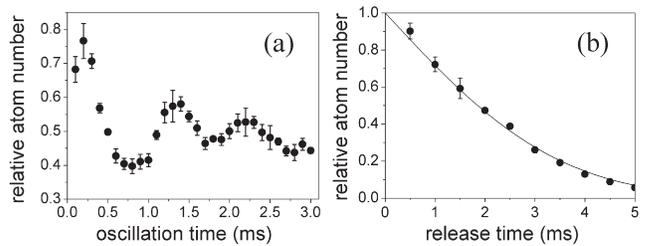}
\caption{\label{freqtemp} (a) Measurement of the horizontal trap
frequency by turning off the Nd:YAG beam for two 2-ms time
intervals separated by a variable oscillation time
\cite{Engler2000a}. The oscillations show $2 \omega_{\rm r}
\approx 2\pi \times 1\,$kHz. (b) Release-and-recapture measurement
of the temperaure by turning off the EW for a variable release
time. The measured temperature is $T=(2.9\pm0.1)\,\mu$K. In both
(a) and (b) the atom number is normalized to the number measured
without interruption of the trapping fields.}
\end{figure}

The effective volume $V_0$ occupied in the microtrap depends on
the horizontal trap frequency $\omega_{\rm r}$ and the temperature
$T$. Fig.~\ref{freqtemp} shows measurements of these two
quantities. In (a) the calculated value of $\omega_{\rm r}/2\pi =
540\,$Hz is nicely confirmed. In (b) the temperature measurement
shows $T=2.9\,\mu$K, obtained at $\Delta \nu_{\rm EW} = 160\,$GHz.
These measurements yield a horizontal $1/e$-radius $\rho_0 =
6.1\,\mu$m, a $1/e$-height $z_0 = 19\,\mu$m and an effective
volume of $V_0 = 2200\mu$m$^3$. With $N = 1.5\times10^5$ atoms a
peak density as high as $7\times10^{13}$cm$^{-3}$ is obtained.
This number is about 300 times higher as compared to our reservoir
and thus the largest gain realized by loading a ``dimple''
\cite{Stamper-Kurn1998c}. The attained density also clearly
exceeds the maximum values reported for Cs in previous dipole
trapping experiments \cite{Boiron1998a, Vuletic1998a, Han2001a}.
The peak phase-space density for a fully unpolarized sample in
$F=3$ is calculated to $7\times10^{-3}$. The very high elastic
collison rate in the dense sample exceeds 2\,kHz and a posteriori
justifies the assumption of a thermal equilibrium distribution.

\begin{figure}
\includegraphics[width=7cm]{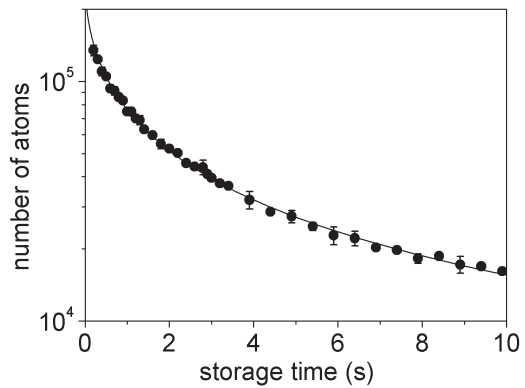}
\caption{\label{loss} Decay of the Cs sample in the surface
microtrap after removing the reservoir at $t=0$. The solid line is
a fit to the data for $\alpha = 0$ that yields $\beta = 4.6 \times
10^{-14}$cm$^3/$s and $K_3 = 5.9 \times 10^{-28}$cm$^6/$s.}
\end{figure}

In order to investigate trap loss at these high densities we have
measured the number of stored atoms as a function of time. The
corresponding experimental data are shown in Fig.~\ref{loss}. Trap
loss can be described by $\dot{N}/N = - \alpha - \beta \langle n
\rangle - K_3 \langle n^2 \rangle$, where $\alpha$ is a
density-independent loss rate and the coefficients $\beta$ and
$K_3$ characterize two-body and three-body inelastic trap loss.
For our trap the mean density and mean quadratic density can be
written as $\langle n \rangle = N/(4V_0)$ and $\langle n^2 \rangle
= N^2/(9V^2_0)$, so that the loss equation yields a differential
equation for $N(t)$.
An attempt to fit the observed decay with $\alpha$, $\beta$, and
$K_3$ as free parametes in addition to the initial particle number
$N(t=0)$ gives ambiguous results. We therefore vary $\alpha$ as an
external parameter in a reasonable range ($\alpha \le
0.1\,$s$^{-1}$) and obtain good fit results for $\beta$ and $K_3$
(see solid line in Fig.~\ref{loss}). In this way we obtain a
two-body loss coefficient of $\sim 5 \times 10^{-14}$cm$^3/$s
together with an upper bound for the three-body coefficient of
$K_3 \lesssim 3 \times 10^{-27}$cm$^6/$s. According to previous
experiments \cite{Hammes2000a} the value for $\beta$ indeed
corresponds to an expected loss coefficient for light-induced
inelastic collisions in the field of the evanescent wave involving
the excitation of repulsive molecular states. The upper bound for
three-body loss appears to be surprisingly low regarding the
resonant scattering properties of Cs \cite{Kerman2001a}.

\begin{figure}
\includegraphics[width=7cm]{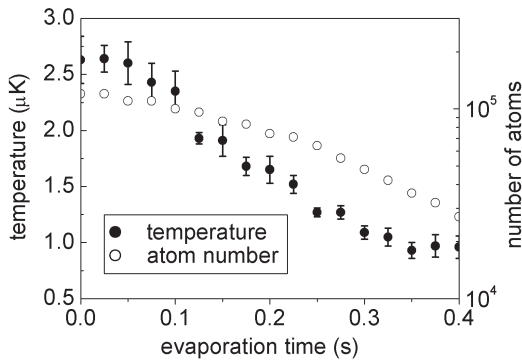}
\caption{\label{evapcooling} Evolution of trapped atom number and
temperature during a 400-ms evaporation ramp, which exponentially
reduces the Nd:YAG beam intensity by a factor of ten.}
\end{figure}

In another series of experiments, we have tried to further
increase the phase-space density by evaporative cooling. For this
purpose, we have lowered the intensity of the Nd:YAG beam in an
exponential ramp with variable time constants. An optimum result,
obtained with a 400-ms ramp down to 10\% of the initial intensity
is shown in Fig.~\ref{evapcooling}. After 350ms of the ramp, with
a potential depth reduced to $\sim$7$\mu$K the temperature reaches
values slightly below 1\,$\mu$K as a result of combined adiabatic
and evaporative cooling. With about $4\times 10^4$ atoms
remaining, the phase-space density of the unpolarized gas in its
seven magnetic substates reaches values slightly above $10^{-2}$.
Further reduction of the intensity leads to increased loss, but
not to lower temperatures. This indicates that evaporation under
our conditions is limited by inelastic loss, which is being
studied in more detail in ongoing experiments.

A great improvement of evaporative cooling in our microtrap can be
expected with polarized atoms in the lowest internal state $F=3,
m_F=3$ at appropriate values of a magnetic field. A factor of
seven will be gained in phase-space density at the same number
density and temperature. Inelastic collisions involving changes of
the magnetic quantum numbers and the release of Zeeman energy will
be energetically suppressed. Moreover, the scattering length can
be tuned into a positive region for magnetic fields larger than
1.7\,mT \cite{Kerman2001a}. In this situation, a further factor of
about 30 in phase-space density and the attainment of Cs-BEC in an
optical surface trap seems feasible with an appropriate
evaporation strategy. A very interesting further prospect would
then be to load the dense gas into a two-dimensional trap formed
by two evanescent waves \cite{Ovchinnikov1991a} and to study the
properties of a two-dimensional surface gas of Cs atoms with their
large and tunable interactions.

In summary, by loading an optical surface microtrap through
elastic collisions from a reservoir we have locally increased the
number and phase-space density by a factor of up to 300. This
demonstrates a very efficient and universal loading scheme for
nondissipative microtraps or future nanotrapping schemes at
surfaces. Moreover it opens up new possibilities to create a
degenerate gas of Cs atoms and to create a two-dimensional surface
quantum gas with tunable interactions.

\begin{acknowledgments}
We gratefully acknowledge support by the Austrian Science Fund
(FWF) within project P15115 and SFB\,15.
\end{acknowledgments}


\end{document}